\documentclass[runningheads]{llncs}
\usepackage[pdftex]{graphicx}
\usepackage{fancybox}
\usepackage{url}
\usepackage{hyperref}

\usepackage{times}
\usepackage{balance}
\usepackage{graphicx}
\usepackage{url}
\usepackage{xcolor}
\usepackage{fontawesome5}

\definecolor{orcidlogocol}{rgb}{0.65, 0.807, 0.223}
\newcommand{\orcid}[1]{$\,$\href{https://orcid.org/#1}{\textcolor{orcidlogocol}{\faOrcid}}}

\begin{document}
\mainmatter              
\title{Deep Learning-Based MR Image Re-parameterization}
\titlerunning{Deep Learning-Based MR Image Re-parameterization}  
%
\author{{Abhijeet Narang\inst{1} \orcid{0000-0003-0677-910X}} \and Abhigyan Raj\inst{1} \and Mihaela Pop\inst{2} \and \\ {Mehran Ebrahimi\inst{3} \orcid{0000-0002-3980-9582}}}
\authorrunning{Abhijeet Narang et al.} 
%
\tocauthor{Abhijeet Narang, Abhigyan Raj, Mihaela Pop, Mehran Ebrahimi}
\institute{Indian Institute of Technology (ISM), Dhanbad, India,\\
\email{\{abhijeetnarang.nr, abhigyanraj\}@gmail.com}
\and
Sunnybrook Research Institute, Toronto, Ontario, Canada,\\
\email{mihaela.pop@utoronto.ca}
\and
Faculty of Science, Ontario Tech University, Oshawa, Ontario, Canada,
\email{mehran.ebrahimi@OntarioTechU.ca}
}

\maketitle              

\begin{abstract}
Magnetic resonance (MR) image re-parameterization refers to the process of generating via simulations of an MR image with a new set of MRI scanning parameters. Different parameter values generate distinct contrast between different tissues, helping identify pathologic tissue. Typically, more than one scan is required for diagnosis; however, acquiring repeated scans can be costly, time-consuming, and difficult for patients. Thus, using MR image re-parameterization to predict and estimate the contrast in these imaging scans can be an effective alternative. In this work, we propose a novel deep learning (DL) based convolutional model for MRI re-parameterization. Based on our preliminary results, DL-based techniques hold the potential to learn the non-linearities that govern the re-parameterization.
\end{abstract}
\section{Introduction}
Current MR imaging techniques \cite{Mansfield_1988,mansfield1982nmr}, provide us with the ability to produce tissue and organ scans with a large set of parameters including: Inversion Time (TI), Repetition Time (TR), Echo Time (TE), etc. Different choices of the scan parameters can generate new images with distinct local tissue contrast properties. Often, the optimal MR parameter setting for an imaging scan is case-specific and not necessarily known in advance \cite{maitra}. Thus, it is desirable to develop models that acquire scans on fixed-parameter sets, and then generate scans corresponding to any parameter setting efficiently. This has motivated the development of several Synthetic MR Imaging methods, such as those exemplified in \cite{ortendahl1984signal,bobman1985synthesized,bobman1986pulse}. MR Image reparameterization refers to generating a given MRI scan with a new set of parameters. Acquiring new MRI scans with new acquisition parameters is costly, time-consuming, and difficult for many patients; thus, MRI re-parameterization to estimate an image with a new set of parameters can be an inexpensive alternative.

While most current MRI simulators \cite{7676360} rely on complex biophysical models to simulate nonlinearities, recent advances in deep learning techniques in different scientific areas have motivated us to develop a DL-based model for MR image re-parameterization

In our work, we propose a coarse-to-fine fully convolutional network for MR image re-parameterization mainly for Repetition Time (TR) and Echo Time (TE) parameters. As the model is coarse-to-fine, we use image features extracted from an image reconstruction auto-encoder as input instead of directly using the raw image. This technique makes the proposed model more robust to a potential overfitting. Based on our preliminary experiments, DL-based methods hold the potential to simulate MRI scans with a new set of parameters. Our deep learning model also performs the task considerably faster than simple biophysical models. To generate our data, we rely on MRiLab \cite{7676360} which is a conventional MR image simulator. Source code is publicly available at \url{https://github.com/Abhijeet8901/Deep-Learning-Based-MR-Image-Re-parameterization }.
\section{Related Works}
While many DL-based techniques have addressed image-to-image translation \cite{isola2017image}, image colorization \cite{nazeri2018image} and super-resolution \cite{nazeri2019edge} problems, to the best of our knowledge there has not been any research on DL-based methods for MR image re-parameterization. 

Previous works have explored DL-based generation of synthetic MR images \cite{fujita2020deep} or used a GAN-based synthesis of MR scans in a given set of parameters \cite{guibas2017synthetic}. However, these used a random latent vector as input, producing a random MR scan for that given set of parameters.

In contrast, recent works employ advanced platforms such as MRiLab \cite{7676360} and Brainweb \cite{cocosco1997brainweb}, which rely on biophysical models that use complex non-linearities to estimate MR images in different parameters. MRiLab is an MR image simulator equipped with the generalized multi-pool exchange model for accurate MRI simulations. 
These works also utilize the multi-pool modeling capabilities of MRiLab to simulate the effects of fat-water interference in macromolecular-rich tissues and validate them in a physical phantom. Brainweb is a Simulated Brain Database generated using an MRI simulator, developed at the McConnell Brain Imaging Centre. This simulator uses first-principles modeling based on the Bloch equations to implement a discrete-event simulation of NMR signal production and realistically models noise and partial volume effects of the image production process. Building these simulators requires physical modeling of MR imaging. Our work in contrast explores the DL-based method to learn these non-linearities that govern the re-parameterization of MR scans from one parameter to another parameter of our choice.
\section{Methodology}
Our method consists of two technical modules: 1) an image reconstruction auto-encoder \cite{10.5555/104279.104293} to extract image features of the input image; and 2) a coarse-to-fine fully convolutional network which utilizes the image features from the auto-encoder and enforces the construction of output image with desired parameters. In the following sections, we introduce how each module works, along with the training and implementation details and the dataset used.

\begin{figure*}[htp]
    \centering
    \includegraphics[width=12.2cm]{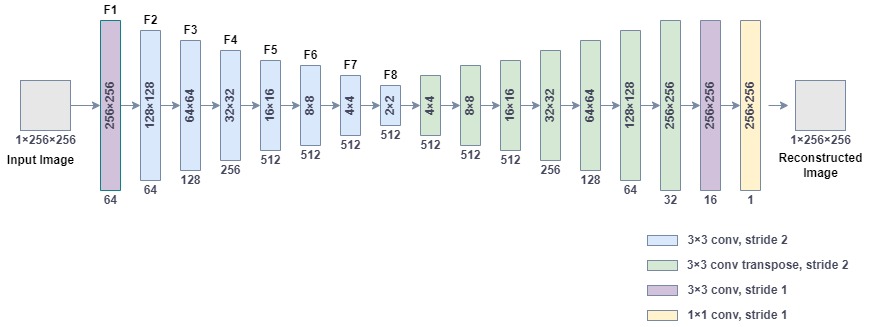}
    \caption{Autoencoder}
    \label{fig:Autoencoder}
\end{figure*}

\begin{figure*}[htp]
    \centering
    \includegraphics[width=12.2cm]{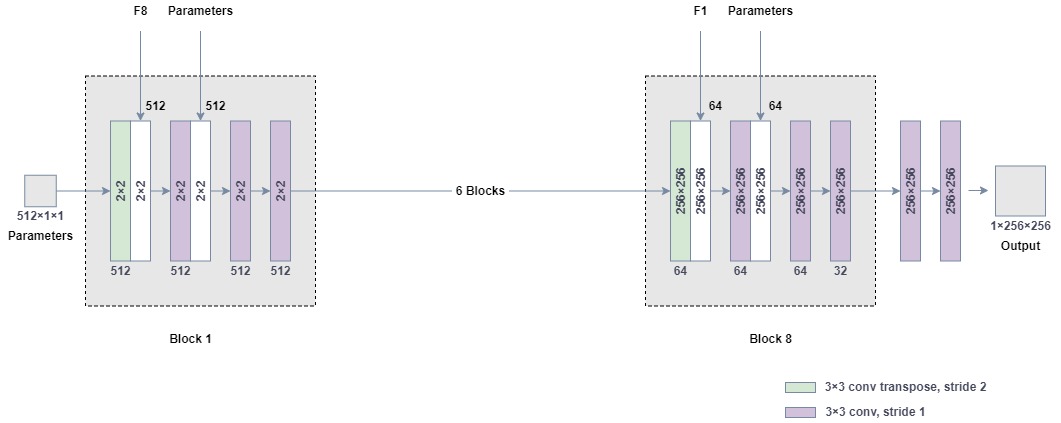}
    \caption{Param-Net}
    \label{fig:Param-net}
\end{figure*}

\subsection{AutoEncoder: Feature Extraction}
An autoencoder is an unsupervised artificial neural network which can learn how to efficiently encode data and then to reconstruct the data back from its encoding. It comprises of two parts – an encoder that progressively downsamples the input using a series of contractive encoding layers and a decoder that reconstructs the input using a series of expansive decoding layers.

Our encoder consists of $8$, $3\times3$ convolutional layers with stride of $2$ for downsampling, each of which is followed by batch normalization layer \cite{43442} and Leaky-ReLU \cite{Maas13rectifiernonlinearities} activation function with a slope of $0.2$. Our decoder consists of $7$, $3\times3$ transposed convolutional layers with a stride of $2$ for upsampling, each followed by batch normalization and ReLU activation function.  This is followed by a $3\times3$ convolutional layer with a stride of $1$ and a $1\times1$ convolutional layer with a stride of $1$ which acts as a pooling layer across the channels. We use $\tanh$ activation function for the last layer as illustrated in Figure \ref{fig:Autoencoder}.

After training the autoencoder, the outputs of $8$ encoding layers are used as input image features for later part of our network. The encoding layers transform the input image into low dimensional vector space. This compression removes redundancy in the input. Providing these low dimensional representations of the input image to our model (i.e., Param-Net instead of directly providing the input image) helps in regularization and makes the learning easier for Param-Net. We notice how this technique is more robust to overfitting and can be generalized to other image-to-image translation tasks.

\subsection{Param-Net}
The param-net is a coarse-to-fine model which uses a series of expansive layers to construct the output image from input image features along with the parameters. It consists of eight blocks where each block consists of a transposed convolutional layer for upsampling followed by three convolutional layers that acts as cross channel parametric pooling layers. The transposed convolutional layer has a filter size of $3\times3$ with a stride of $2$ for upsampling. The size of the output of transposed convolutional filter of block $i$, $1\leq i \leq 8$ will match with the size of the output of $9-i$ layer of the auto-encoder. We stacked the image features from that layer of the auto-encoder with the output of the third transposed convolutional layer. We passed this through a $3\times3$ convolutional layer with a stride of $1$. We then stack the parameters with the output of this convolutional layer. This is then passed through two $3\times3$ convolutional layers with a stride of 1 which also acts as a cross channel parametric pooling layer. Each of these three convolutional layers is followed by instance normalization and \cite{Ulyanov_Instance} Leaky ReLU activation function with the slope of 0.2. The last block is followed by two $3\times3$ convolutional layers with a stride of $1$. This is followed by tanh activation function which produces the output image.

The parameters stacked in between the layers are shaped in the size of the layer it is being stacked to. The reason we pass the parameters in each block is that if we only passed them in the first block, it would be difficult for the later blocks to retain them. This problem is somewhat similar to the degradation problem faced by deep networks such as ResNet \cite{He_ResNet} which is solved by skip connections that allows the later layers to access features from the previous layers directly as illustrated in Figure \ref{fig:Param-net}.

We use this architecture for two models: 1) Default-to-Param 2) Param-to-Param as explained below.

\subsubsection{Default-to-Param}
In this model, the input image is always of a fixed parameter value. We chose those default values to be, TE = 50ms, TR = 4.5s. For this model, we only pass the parameters of the desired output image to the Param-Net.
\subsubsection{Param-to-Param}
In this model, the parameters of the input image are not fixed. Hence, we also pass the input image parameters along with the output image parameters. This model is more generalizable than the previous one but also has a more complex non-linearity to learn and hence, lags behind in performance compared to default-to-param. Hence there is a trade-off between both the models.

\subsection{Dataset}
For our training, we require the MRI scans in two different parameter settings of \{TE, TR\}. One serves as input to the model, and the other as the ground truth corresponding to the desired parameter setting to compute the loss. We use MRiLab \cite{7676360} which is an MRI Simulator to generate these synthetic brain scans in different parameter settings of \{TE, TR\}. We generated these brain MRI scans for 200 random pairs of \{TE, TR\}. The TR values were chosen uniformly at random in the range 1.2 s to 10s. The TE values ranged from 20 ms to 1s non-uniformly. The distribution was such that lower TE values were selected with higher probability. This was done because the scans were more sensitive toward changes in lower values of TE. The T1 and T2 relaxation times used by MRiLab were matrices of size $108\times 90\times 90$ with values in the range 0s to 4.5s for T1 and 0s to 2.2s for T2. For each pair of \{TE, TR\}, we generated 24 different 2D axial MR slices of a 3D brain volume, so in total we obtained 4800 MR slices. We used 1500 samples of these slices for training, while the rest were kept for testing. The generated scans were rescaled to a $256\times 256$ matrix.

\subsection{Network Training}
The training process comprises of two sequential phases: training of the auto-encoder, and training of the Param-Net. For the first phase, the auto-encoder was trained on a subset of Places-365 dataset \cite{zhou2017places} and fine-tuned using MR Image dataset. For the second phase, we train the Param-Net on MRiLab dataset. The weights of the auto-encoder are frozen during this phase. For both phases, we used Adam optimizer \cite{Kingma_Adam} and weight initialization as proposed by \cite{Glorot_Xavier}. Mean squared error (MSE) loss was used for both phases.
\section{Results}
To evaluate the performance of our model, we employ Peak Signal-to-Noise Ratio (PSNR) and Mean Absolute Error (MAE).

\subsection{MRiLab Testset}
\begin{figure*}[htp]
    \textbf{\large A. MRiLab Dataset Results}
    
    \vspace{10mm}
    
    \begin{center}
    \includegraphics[width=12.119cm]{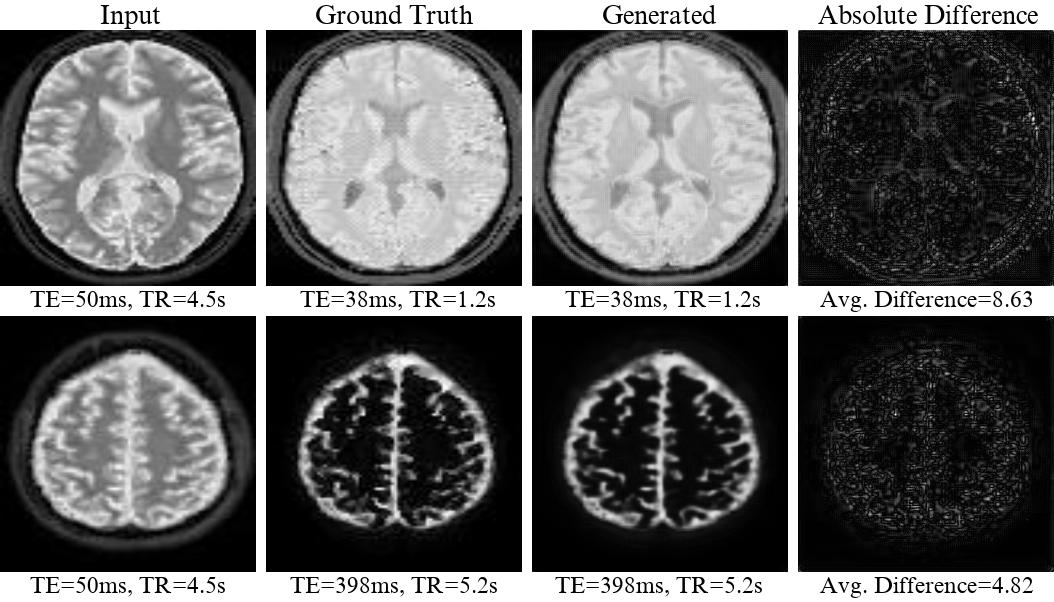}
    \caption{Re-parameterization results for Default-to-Param Model, signal intensity (SI) and the absolute difference are represented in arbitrary units.}
    \label{fig:MRiLab Default to Param}
    
    \vspace{10mm}
    
    \includegraphics[width=12.119cm]{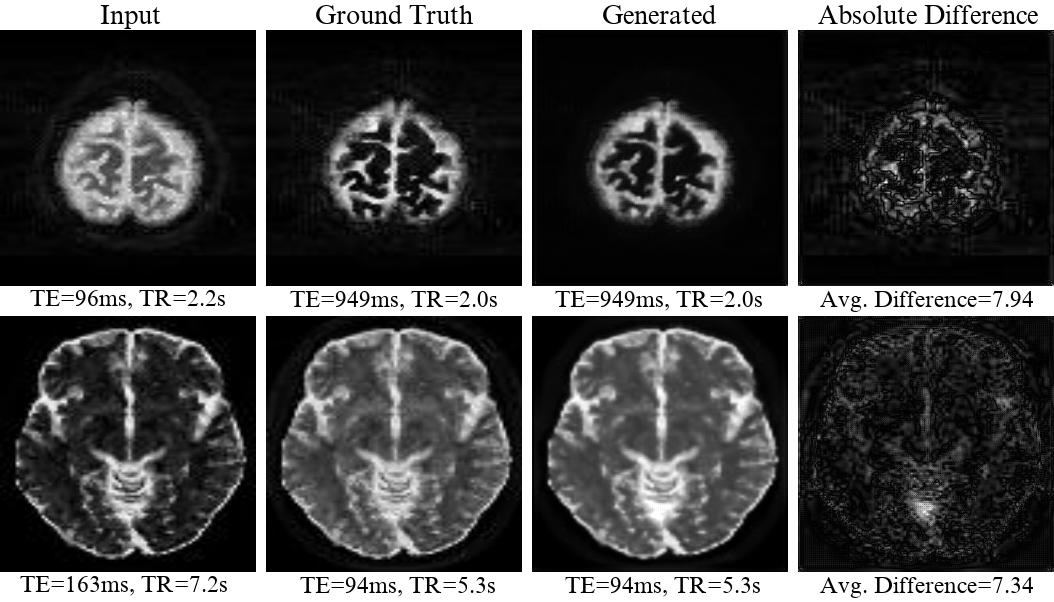}
    \caption{Re-parameterization results for Param-to-Param Model, signal intensity (SI) and the absolute difference are represented in arbitrary units.}
    \label{fig:MRiLab Param to Param}
    \end{center}
\end{figure*}

Figures \ref{fig:MRiLab Default to Param} and \ref{fig:MRiLab Param to Param} show examples of axial slices from our testset MRiLab for both models. The figures show the input, ground truth, model's prediction and the absolute difference between the ground truth and the prediction. For all images the signal intensity (SI) and absolute difference are represented in arbitrary units. The evaluation metrics on the test-set of 3300 images from MRiLab can be seen in Table \ref{MriLab TestSet Results}.

\begin{table*}[ht]
\caption{Performance of both models on MRiLab Testset}
\label{MriLab TestSet Results}
\centering
\begin{tabular}{|c||p{15mm}|p{15mm}|p{15mm}|p{15mm}|}
\hline
Model Type & Mean PSNR & Std. Dev. of PSNR & Mean \newline Absolute Difference  & Std. Dev. of \newline Absolute Difference \\
 
\hline
Default-to-Param & 30.34 & 1.04 & 5.25 & 1.74 \\
\hline
Param-to-Param & 24.80 & 1.68 & 9.51 & 4.55 \\
\hline
\end{tabular}
\end{table*}

\subsection{Brainweb Testset}
\begin{figure*}[htp]
    \textbf{\large B. BrainWeb Dataset Results}
    
    \vspace{10mm}
    
    \begin{center}
    \includegraphics[width=12.119cm]{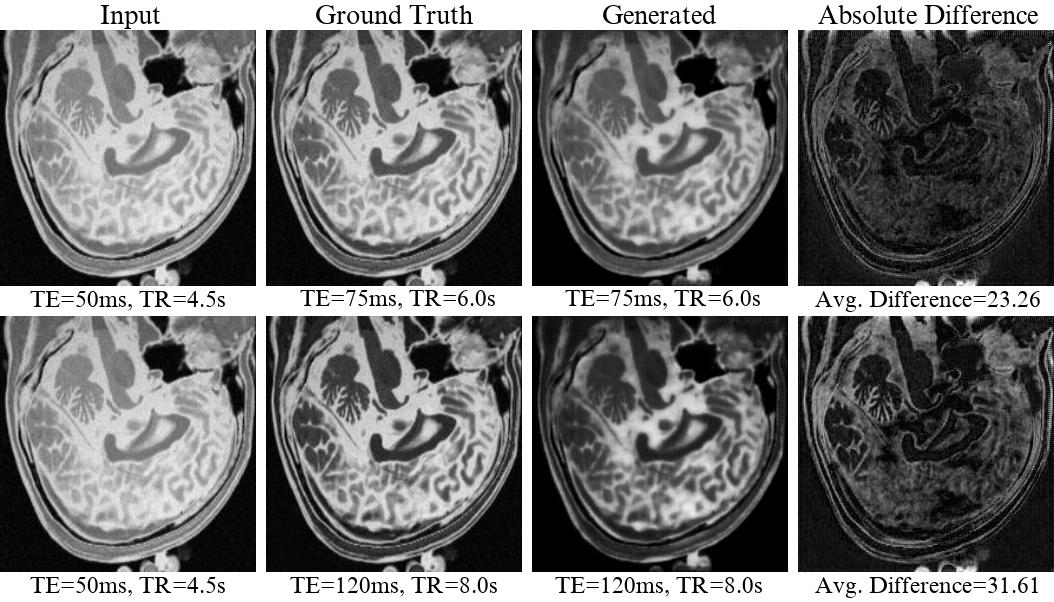}
    \caption{Re-parameterization results for Default-to-Param Model, signal intensity (SI) and the absolute difference are represented in arbitrary units.}
    \label{fig:BrainWeb Default to Param}
    
    \vspace{10mm}
    
    \includegraphics[width=12.119cm]{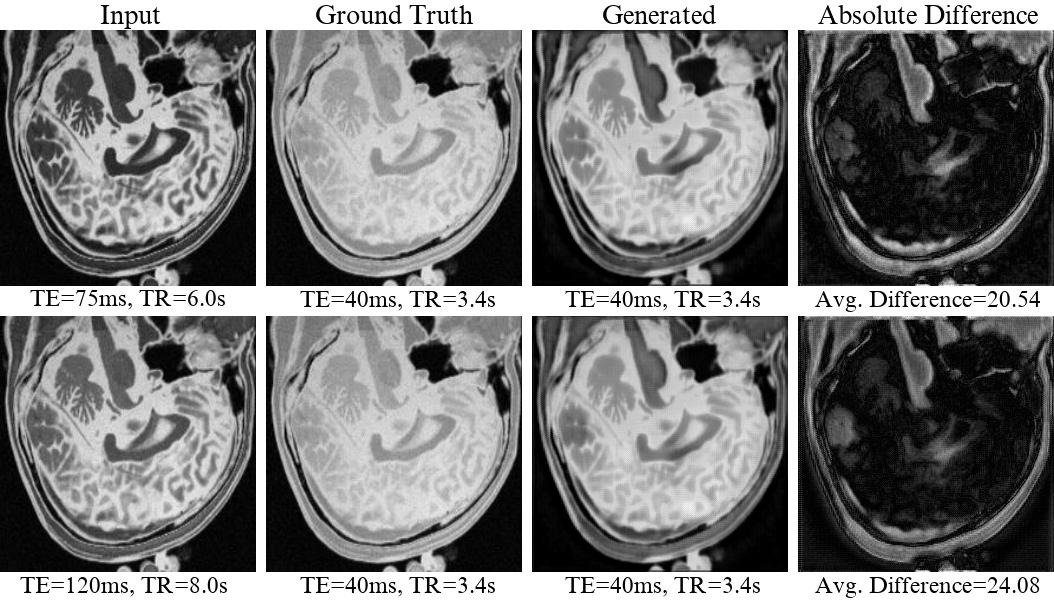}
    \caption{Re-parameterization results for Param-to-Param Model, signal intensity (SI) and the absolute difference are represented in arbitrary units.}
    \label{fig:BrainWeb Param to Param}
    \end{center}
\end{figure*}

Brainweb is a simulated brain database that contains a set of realistic MRI data volumes produced by an MRI Simulator. We used this tool to generate test scans in 5 different parameter settings. The results can be seen in Figures \ref{fig:BrainWeb Default to Param} and \ref{fig:BrainWeb Param to Param} for both models. The evaluation metrics on this test-set can be found in Table \ref{BrainWeb TestSet Results}.

\begin{table*}[ht]
\caption{Performance of both models on Brainweb Testset}
\label{BrainWeb TestSet Results}
\centering
\begin{tabular}{|c||p{15mm}|p{15mm}|p{15mm}|p{15mm}|}
\hline
Model Type & Mean PSNR & Std. Dev. of PSNR & Mean \newline Absolute Difference  & Std. Dev. of \newline Absolute Difference \\
 
\hline
Default-to-Param & 20.12 & 2.89 & 22.3 & 6.73 \\
\hline
Param-to-Param & 16.12 & 2.21 & 34.09 & 9.98 \\
\hline
\end{tabular}
\end{table*}

It can be observed that the default-to-param model achieves a higher mean PSNR and lower MAE compared to Param-to-Param model. We believe this is due to the more complex nonlinearity associated with the task of reparameterizing from any parameter than from a fixed parameter.
\section{Conclusion}
In summary, our simulation study showed that DL-based methods can be used for MR image re-parameterization. Based on our preliminary results, we suggest that DL-based methods hold the potential to generate via simulations MR imaging scans with a new set of parameters.
Future work can focus on varying larger number of acquisition parameters. This approach could also be utilized for T1/T2 mapping, based on the availability of sufficient training data.

\section*{Acknowledgements}
This work was supported in part by the Natural Sciences and Engineering Research Council of Canada (NSERC).

%

%
\newpage

\newpage
\end{document}